\documentclass[12pt,letterpaper,onecolumn]{article}
\usepackage{cite,graphicx,subfigure,amssymb,color,epsfig,amsmath,times,setspace,url}

\usepackage{todonotes}

\makeatletter
  \renewcommand\@biblabel[1]{#1.}     
  \def\@cite#1#2{({#1\if@tempswa , #2\fi})}
  \def\eqref#1{[\ref{#1}]}
\makeatother
\makeatletter
  \def\tagform@#1{\maketag@@@{[#1]\@@italiccorr}}
\makeatother

\begin{document}

\begin{center}
{\huge \bf Fast Sub-millimeter Diffusion MRI using gSlider-SMS and SNR-Enhancing Joint Reconstruction}
\end{center}
 
\vfill

\noindent Justin P. Haldar$^{1}$, Qiuyun Fan$^{2}$, Kawin Setsompop$^{2}$

\vfill

$^1$Ming Hsieh Department of Electrical and Computer Engineering, University of Southern California, Los Angeles, CA, 90089, USA

$^2$Athinoula A. Martinos Center for Biomedical Imaging, Charlestown, MA, 02129, USA.

\vfill 

\begin{singlespace}
\noindent {\bf Address Correspondence to:}
\vspace{.25in}

\noindent Justin P. Haldar

\noindent University of Southern California

\noindent University Park Campus

\noindent 3740 McClintock Avenue

\noindent Hughes Aircraft Electrical Engineering Center (EEB) \#442, M/C 2564

\noindent Los Angeles, CA 90089

\noindent Tel: (213) 740-2358

\noindent Email: jhaldar@usc.edu
\end{singlespace}

\vfill

\noindent MANUSCRIPT BODY APPROXIMATE WORD COUNT = $\sim$5000

\newpage

\section*{ABSTRACT}

\noindent {\bf Purpose:} 

We evaluate a new approach for achieving diffusion MRI data with high spatial resolution, large volume coverage, and fast acquisition speed.

\noindent {\bf Theory and Methods:}  

A recent method called gSlider-SMS enables whole-brain sub-millimeter diffusion MRI with high signal-to-noise ratio (SNR) efficiency.  However, despite the efficient acquisition, the resulting images can still suffer from low SNR due to the small size of the imaging voxels.  This work proposes to mitigate the SNR problem by combining gSlider-SMS with a regularized SNR-enhancing reconstruction approach.

\noindent {\bf Results:}

Illustrative results show that, from gSlider-SMS data acquired over a span of only 25 minutes on a 3T scanner, the proposed method is able to produce 71  MRI images (64 diffusion encoding orientations with $b=$1500 s/mm$^2$, and 7 images without diffusion weighting) of the entire \emph{in vivo} human brain  with nominal 0.66 mm spatial resolution.  Using data acquired from 75 minutes of acquisition as a gold standard reference, we demonstrate that the proposed SNR-ehancement procedure leads to substantial improvements in estimated diffusion parameters compared to conventional gSlider reconstruction.  Results also demonstrate that the proposed method has advantages relative to denoising methods based on low-rank matrix modeling.  A theoretical analysis of the trade-off between spatial resolution and SNR suggests that the proposed approach has high efficiency.

\noindent {\bf Conclusion:}

The combination of gSlider-SMS with advanced regularized reconstruction enables high-resolution quantitative diffusion MRI from a relatively fast acquisition.

\section*{KEYWORDS}
Diffusion MRI; High Resolution; Fast Imaging; Constrained Reconstruction; Denoising;

\newpage

\section{Introduction}

It is challenging to acquire quantitative whole-brain \emph{in vivo} human diffusion MRI data with sub-millimeter resolution. One of the main obstacles is the limited signal-to-noise ratio (SNR) associated with small voxel sizes.  For illustration,  a classical theoretical analysis \cite{macovski1996} suggests that if we want the SNR of a 660$\mu$m acquisition to match the SNR of a 2mm acquisition with all other imaging parameters held equal, then the 660$\mu$m data would need over 770$\times$ more data averaging,  because SNR is proportional to the product between the voxel volume and the square-root of the number of averages.  This degree of averaging is not practical for \emph{in vivo} human applications.

Nevertheless, there has been considerable recent progress towards achieving higher-resolution diffusion MRI  \cite{holdsworth2019}.  On the data-acquisition side, some of the major advances have been driven by the use of simultaneous multi-slice (SMS) \cite{setsompop2012,setsompop2013,sotiropoulos2013} and volumetric \cite{jeong2003,engstrom2013,nguyen2014,frost2014,chang2015,van2015,holtrop2016} excitation methods, which yield substantially higher SNR efficiency relative to standard 2D spatial encoding methods.  Related approaches acquire multiple sets of low-reslution high-SNR images with RF encoding, and computationally fuse these low-resolution images together into a ``super-resolved'' high resolution image \cite{greenspan2009,scherrer2012,vansteenkiste2016,ning2016,hennel2018,setsompop2018,wang2018,liao2018}.   One recent approach, gSlider-SMS \cite{setsompop2018,wang2018,liao2018}, combines  SMS and RF encoding ideas, and is able to achieve whole-brain \emph{in vivo} human diffusion MRI with isotropic sub-millimeter resolution with relatively high SNR-efficiency.  Unfortunately, the small size of the imaging voxels still means that gSlider-SMS often requires signal averaging.

Substantial progress has also been made in the use of signal processing methods to improve the SNR of noisy diffusion images (for example, see the many denoising techniques that are reviewed in Ref.~\cite{haldar2012}, as well as more recent methods such as Refs.~\cite{lam2013,varadarajan2015,veraart2016,sperl2017,kafali2018}).  One such approach, SNR-enhancing joint reconstruction (SER) \cite{haldar2008a,haldar2012,kim2015a,haldar2011c}, is designed to spatially smooth the data to improve SNR while also preserving image edge features that are shared between different DWIs to avoid the loss of important high-resolution information.  In contrast to many other denoising techniques that are difficult to characterize theoretically, the SER approach has a strong theoretical characterization that can provide the user with a precise understanding of the trade-offs associated with SER.  It is known that SER is associated with a controllable trade-off between SNR improvement and spatial resolution \cite{haldar2012,haldar2008a,haldar2011c,kim2015a}, and the SER theory empowers the user to choose an appropriate balance between these two factors.   Importantly, it has been shown that significant gains in SNR can be achieved using spatial smoothing with only modest corresponding losses in spatial resolution  \cite{haldar2012,haldar2008,haldar2011c,kim2015a,haldar2009c,haldar2008a,haldar2011a}, and that the impact of these modest resolution losses can be largely mitigated because of the edge preservation properties of SER. In addition, previous analyses \cite{haldar2012,haldar2011c} suggest that the regularization penalty behaves in a stable way in the presence of assumption violations, which can provide more confidence in its good performance when applied to data with non-standard features.  For example, previous work has shown that SER works well with images of both normal  and injured tissue \cite{kim2015a,haldar2012} (as would be predicted by theory), while other popular nonlinear denoising approaches may work well with normal tissue but can yield problematic results when subtle injuries are present \cite{kim2015a}.

In this work, we propose and study the combination of  gSlider-SMS  with SER, and show that this combination enables state-of-the-art performance in achieving fast sub-millimeter diffusion MRI.  Preliminary accounts of portions of this work were previously presented in Ref.~\cite{haldar2016}.

\section{Theory and Methods}
\subsection{gSlider-SMS}
For simplicity and brevity, this paper will describe gSlider-SMS \cite{setsompop2018,wang2018,liao2018} for the simplified case in which parallel imaging and SMS reconstruction have already been performed,  and it remains to reconstruct the high-resolution spatial information along the slice dimension.   Rather than exciting thin slices, conventional gSlider encoding excites slabs that are several times thicker than the desired high-resolution slice thickness, and uses multiple RF slab-encodings across consecutive TRs to resolve the constituent sub-slices that comprise the slab. For the purposes of our description and without loss of generality, we will assume that the thick slab is five times larger than the nominal slice thickness and that we seek to recover five constituent sub-slices.    To achieve this, gSlider acquires a series of different  images of the same slab, where the RF pulse is varied in each repetition to apply a distinct phase modulation pattern  to the sub-slices, in a manner similar to Hadamard encoding and related methods \cite{maudsley1980,oh1984}. 

If $\mathbf{b}$ denotes the vector concatenating the collection of thick-slab images acquired with different RF encoding profiles and $\mathbf{f}$ represents the high-resolution image, then the RF-encoding model (in the absence of motion-induced phase variations) can be written as $\mathbf{b} = \mathbf{A}\mathbf{f}$, where matrix $\mathbf{A}$ captures the RF encoding procedure.  Given this model, it is possible to recover the desired high-resolution slice information through standard linear Tikhonov-regularized reconstruction \cite{setsompop2018}:
\begin{equation}
\hat{\mathbf{f}} = (\mathbf{A}^H\mathbf{A} + \lambda \mathbf{I})^{-1}\mathbf{A}^H \mathbf{b},\label{eq:grecon}
\end{equation}
where $\lambda$ is a regularization parameter and $\mathbf{I}$ is the identity matrix.  In practice, real data will possess motion-induced phase variations, and it is necessary to apply appropriate phase correction to $\mathbf{b}$ prior to applying Eq.~\eqref{eq:grecon}.  We refer to the phase-corrected version of Eq.~\eqref{eq:grecon} as conventional gSlider reconstruction \cite{setsompop2018}.

The gSlider encoding matrix $\mathbf{A}$ is designed to have two useful properties that allow it to outperform alternative RF encoding schemes like Fourier \cite{oh1984}, Hadamard \cite{oh1984}, and random \cite{haldar2010a,kim2015} RF encoding.  First, the gSlider is designed to have high image SNR for all the RF encodings, unlike e.g. Hadamard or Fourier encoding.  This  enables robust phase estimation and correction, which is important due to the phase inconsistencies of diffusion MRI  \cite{setsompop2018}.  Second, the $\mathbf{A}$ matrix is designed to be well conditioned, which ensures that the matrix inversion in Eq.~\eqref{eq:grecon} does not substantially exacerbate noise, and actually leads to substantial SNR gains relative to 2D slice-by-slice acquisition \cite{setsompop2018}.  In addition, the use of RF encoding can offer sharper point-spread functions with less signal leakage/Gibbs ringing than k-space Fourier encoding when used for thin-slab encoding.

\subsection{SNR-Enhancing Joint Reconstruction}

While gSlider encoding is relatively SNR-efficient, the use of small isotropic voxel sizes means that SNR is still a limiting factor.  We propose to use SER \cite{haldar2012,kim2015a,haldar2008a,haldar2011c} to address this issue. SER uses SNR-efficient spatial smoothing to reduce noise perturbations, while leveraging the structural similarity between different diffusion weighted images (DWIs) to avoid blurring high-resolution image features.  Since the  data considered in this work is acquired with a partial Fourier readout, our SER approach also uses phase modeling to prevent the loss of in-plane resolution while also automatically accounting for the phase discrepancies associated with diffusion MRI.

\begin{figure}[ht]
\includegraphics[width=\textwidth]{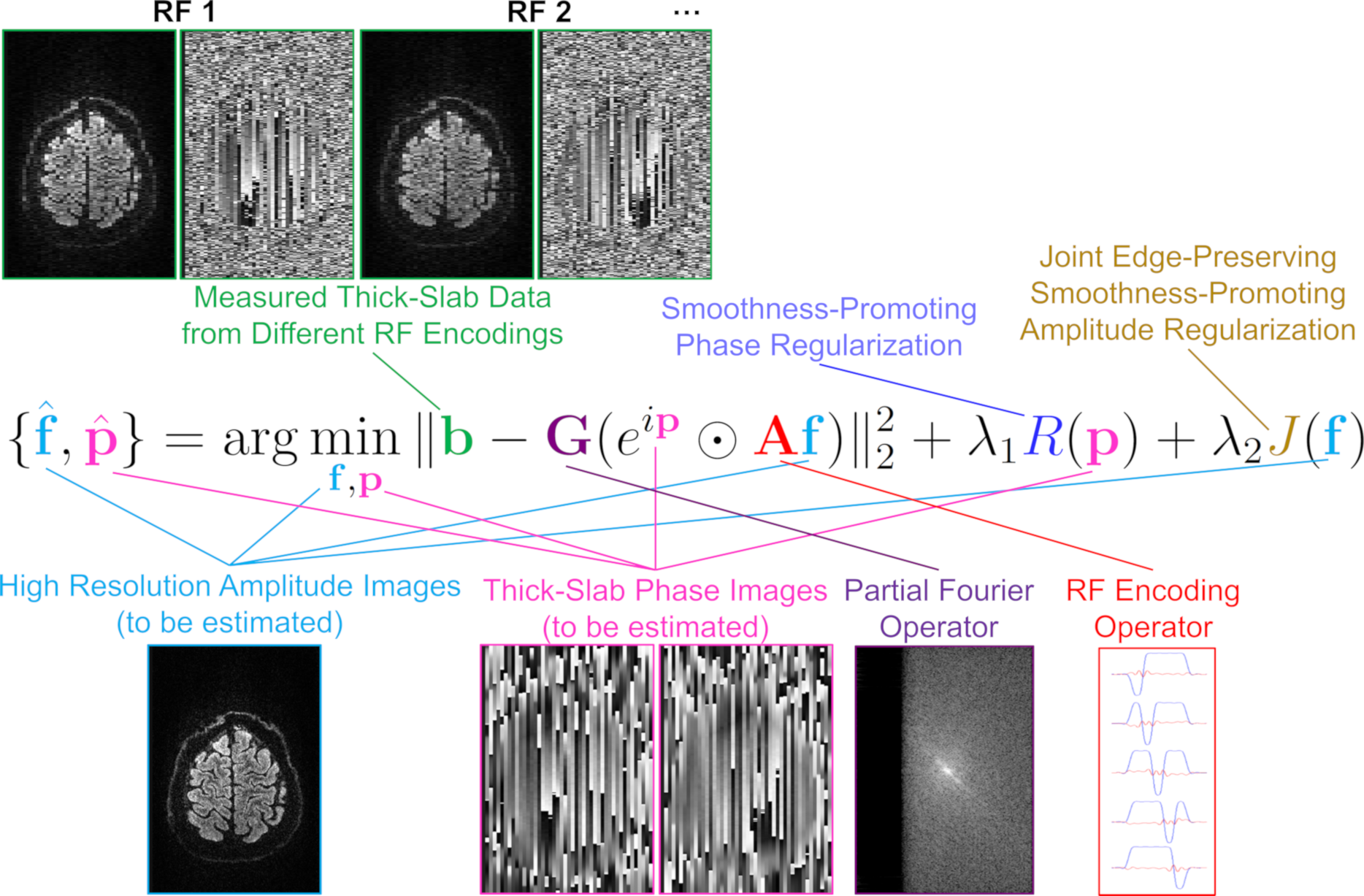}
\caption{A graphical overview of the proposed image reconstruction formulation, depicting all of the relevant inputs, outputs, and operators.  In this figure, RF encoding is performed using sagittal slabs, and the phase smoothness prior is applied in 2D within each of the sagittal slabs.}
\label{fig:overview}
\end{figure}

SER is performed by solving an optimization problem that couples regularized denoising and partial Fourier reconstruction as in Ref.~\cite{haldar2012} (with appropriate modifications to incorporate the gSlider RF encoding model) with regularized phase estimation as in Refs.~\cite{haldar2011d,zhao2012a}.\footnote{Our previous SER work \cite{haldar2012} used a two step approach, in which the phase term for each image was estimated \emph{a priori} using a low-resolution reconstruction of each image, and this phase estimate was subsequently used to estimate the image amplitude.  This approach works well and is still used as an initialization in the present work. However, our new formulation  allows for potentially fixing any  errors that may exist in the initial phase estimate. } Specifically, we solve 
\begin{equation}
\{\hat{\mathbf{f}}, \hat{\mathbf{p}}\} = \arg\min_{\mathbf{f}, \mathbf{p}} \| \mathbf{b} - \mathbf{G}(e^{i\mathbf{p}} \odot \mathbf{A}\mathbf{f}) \|_2^2 + \lambda_1 R(\mathbf{p}) + \lambda_2 J(\mathbf{f}).\label{eq:opt}
\end{equation}
The data vector $\mathbf{b} \in \mathbb{C}^{(5 N_1 N_2 N_s N_d)\times 1}$ contains the complex-valued voxel values of the RF-encoded slab images (obtained after slice-GRAPPA reconstruction).  When specifying the dimension of $\mathbf{b}$, we have assumed that there are 5 RF encodings, that each RF-encoded slab image contains $N_1 \times N_2$ voxels in-plane, that we have acquired slabs at $N_s$ different slab positions to achieve full volume coverage, and that for diffusion encoding, we acquire data for $N_d$ images with varying diffusion parameters.  We use $N_3 = 5 N_s$ to represent the total number of high-resolution thin sub-slices that we wish to recover. The optimization variable $\mathbf{f}\in \mathbb{R}^{(N_1 N_2 N_3 N_d)\times 1}$ is the vector of  the unknown high-resolution (i.e., based on thin sub-slices) real-valued image voxel amplitudes for all of the diffusion weighted images, while $\mathbf{A} \in \mathbb{C}^{(5 N_1 N_2 N_s N_d) \times (N_1 N_2 N_3 N_d)}$ is the matrix modeling the RF-encoding procedure.  The optimization variable $\mathbf{p}\in \mathbb{R}^{(5 N_1 N_2 N_s N_d)\times 1}$ is used to model the unknown phase of each measured thick-slab acquisition (to enable phase correction to compensate for motion-induced phase discrepancies for gSlider-SMS reconstruction, as well as providing the phase constraints needed for resolution recovery from partial Fourier data acquisition), and $\mathbf{G}\in\mathbb{C}^{(5 N_1 N_2 N_s N_d) \times (5 N_1 N_2 N_s N_d)}$ is a matrix used to model the in-plane point-spread function of partial Fourier acquisition (i.e., Fourier transformation of each image into k-space, setting the unmeasured portion of k-space equal to zero, followed by inverse Fourier transformation).  The symbol $\odot$ is used to denote the Hadamard product (i.e., elementwise multiplication of two vectors) and $e^{i \mathbf{p}}\in \mathbb{C}^{(5 N_1 N_2 N_s N_d)\times 1}$ is used to denote the vector of exponentiated phase values.  A graphical depiction of this equation is shown in Fig.~\ref{fig:overview}.

The first term in Eq.~\eqref{eq:opt} is a standard least-squares data consistency penalty that encourages the reconstructed high-resolution image to be consistent with the low-resolution measured data.  Partial Fourier constraints are imposed by requiring that $\mathbf{f}$ is real-valued \cite{haldar2012}.

The second term $R(\mathbf{p})$ in Eq.~\eqref{eq:opt} is a regularization penalty  that encourages the estimated image phase to be smooth within each acquired image slab, but with no constraints on the phase behavior between different slabs or between different diffusion weighted images (as appropriate in the presence of random phase variations).  To avoid difficulties associated with phase-wrapping and the non-uniqueness of phase, the regularization penalty is designed to regularize the exponentiated phase (a choice that is insensitive to 2$\pi$ phase jumps \cite{haldar2011d,zhao2012a}), with 
\begin{equation}
\begin{split}
R(\mathbf{p}) &= \sum_{s=1}^{N_s} \sum_{q=1}^{N_d}\sum_{k = 1}^5 \sum_{n=1}^{N_1 N_2}  \sum_{m \in \Omega_n} |e^{i p_{sqkn}} - e^{i p_{sqkm}}|^2 \\
&= \|\mathbf{D}e^{i\mathbf{p}}\|_2^2.\label{eq:phasereg}
\end{split}
\end{equation}
In this equation, the $p_{sqkn}$ values denote the phase value from $\mathbf{p}$ corresponding to the $n$th voxel in the $s$th slab with the $k$th RF encoding and the $q$th diffusion encoding, $\Omega_n$ corresponds to the set of 4  voxels that are immediate in-plane spatial neighbors to the $n$th voxel, and $\mathbf{D}$ is the matrix representation of the finite differencing operation.   

The third term $J(\mathbf{f})$ in Eq.~\eqref{eq:opt} is a regularization penalty that leverages the prior knowledge that $\mathbf{f}$ is expected to be spatially smooth within each DWI volume, but also has edge structures that are common between different DWIs which ideally should be preserved by the reconstruction procedure \cite{haldar2012,haldar2008a,haldar2011c,kim2015a}. If we use $f_n^q$ to denote the entry from  $\mathbf{f}$ corresponding to the $n$th voxel of the $q$th DWI, then  we choose $J(\mathbf{f})$ following Refs.~\cite{haldar2012,haldar2008a,haldar2011c,kim2015a,lam2013}  as 
\begin{equation}
 J(\mathbf{f}) =  \sum_{n=1}^{N_1 N_2 N_3} \sum_{m \in \Delta_n}  \Psi\left(\sqrt{ \sum_{q=1}^{N_d} | f_n^q - f_m^q |^2} \right),\label{eq:j1}
\end{equation}
where $\Delta_n$ corresponds to the set of 6 voxels that are immediate volumetric spatial neighbors of the $n$th voxel, and $\Psi(\cdot)$ is the convex Huber function:
\begin{equation}
\Psi(t) = \left\{ \begin{array}{ll} t^2, & t \leq \xi \\ 2\xi t - \xi^2, & t >\xi. \end{array} \right.
\end{equation}
The Huber function is well known as a regularization penalty that is both convex and edge-preserving, and converges to a scaled version of the standard $\ell_1$-norm in the limit as $\xi$ approaches zero.  In the limit of small $\xi$,  the penalty function of Eq.~\eqref{eq:j1}  promotes joint sparsity of the image edges \cite{haldar2008a}, and becomes equivalent to a joint sparsity-promoting regularization penalty function that is popular in the multi-contrast compressed sensing MRI literature \cite{trzasko2011,huang2014}.  We prefer to use larger values of $\xi$, since this enables quantitative theoretical characterizations of the resolution and noise characteristics of the reconstructed images while still encouraging shared edge preservation \cite{haldar2012,haldar2011c}.  These theoretical characterizations offer a number of advantages as described previously. 

In this work, we use global rescaling of the images in $\mathbf{b}$ prior to image reconstruction, such that the median voxel intensity has the same magnitude across each of the DWIs.  This normalization is useful to avoid the estimated shared-edge structure from being dominated by the images with the highest intensity (i.e., the unweighted $b=0$ images), and the normalization is easily removed once SER has been completed.  This rescaling is equivalent to the use of  additional hyperparameters that were present in the original formulation of SER \cite{haldar2012}, although leads to simpler equations.

As described in more detail in Ref.~\cite{haldar2012}, we choose the reconstruction parameter $\xi$ to be large enough that the estimated edge structures are relatively free of visible noise influences, and choose $\lambda_2$ based on the level of desired SNR improvement (e.g., in the results we show later, we leverage the theoretical characterizations that are available for SER and choose $\lambda_2$ so that the SNR improvement in most regions of the brain is at least as good as 3$\times$ averaging).  The parameter $\lambda_1$ is chosen heuristically to obtain estimated phase maps that are not too noisy but which are also not visually oversmoothed.

For optimization of Eq.~\eqref{eq:opt}, we choose an iterative alternation-based approach that guarantees monotonic decrease (and, therefore, convergence, although not necessarily to a global optimum) of the cost function value.  In the first stage of the $i$th iteration, we estimate the value of the image magnitude $\hat{\mathbf{f}}$ given the previous estimate of the image phase according to
\begin{equation}
\hat{\mathbf{f}}^{(i)} = \arg\min_{\mathbf{f}} \| \mathbf{b} - \mathbf{G}(e^{i\hat{\mathbf{p}}^{(i-1)}} \odot \mathbf{A}\mathbf{f}) \|_2^2 +  \lambda_2 J(\mathbf{f}),\label{eq:con}
\end{equation}
subject to the constraint that $\mathbf{f}$ be real-valued.  The optimization problem in Eq.~\eqref{eq:con} has identical form to that considered in Ref.~\cite{haldar2012}, and we find a globally optimal solution using the simple iteratively-reweighted least-squares optimization algorithm that was previously proposed for this problem \cite{haldar2012}.

In the second stage of the $i$th iteration, we estimate the value of the image phase $\hat{\mathbf{p}}$ given the  estimate of the image amplitude according to
\begin{equation}
\hat{\mathbf{p}}^{(i)} = \arg \min_{\mathbf{p}} \| \mathbf{b} - \mathbf{G}(e^{i\mathbf{p}} \odot \mathbf{A}\hat{\mathbf{f}}^{(i)}) \|_2^2 + \lambda_1 \|\mathbf{D}e^{i\mathbf{p}}\|_2^2.\label{eq:phi}
\end{equation}
This is a nonlinear optimization problem that has exactly the same structure as the optimizations considered in previous work \cite{haldar2011d,zhao2012a}. Similar to the previous work \cite{haldar2011d,zhao2012a} we use the nonlinear conjugate gradient (NCG) algorithm with analytically-evaluated gradients to find a local minimum. In particular, we use a variation of NCG that uses the Polak-Ribi\`{e}re method constrained by the Fletcher-Reeves method \cite{gilbert1992a}.  The gradient of Eq.~\eqref{eq:phi} can be computed analytically as
\begin{equation}
\nabla f (\mathbf{p}) = 2 \mathrm{imag}\left( e^{-i\mathbf{p}} \odot   \mathbf{A}\hat{\mathbf{f}}^{(i)} \odot (\mathbf{G}^H\mathbf{G}(e^{i\mathbf{p}} \odot \mathbf{A}\hat{\mathbf{f}}^{(i)})-\mathbf{G}^H\mathbf{b})+\lambda_1 e^{-i\mathbf{p}} \odot (\mathbf{D}^H\mathbf{D} e^{i\mathbf{p}}) \right),
\end{equation}
where $\mathrm{imag}(\mathbf{x})$ denotes the imaginary part of a complex vector $\mathbf{x}$, and $^H$ denotes the standard conjugate transpose matrix operation.

The two optimization steps represented by Eqs.~\eqref{eq:con} and \eqref{eq:phi} are iterated until convergence, or until a maximum number of iterations has been met.

\subsection{Data Acquisition and Processing}
Whole-brain gSlider-SMS diffusion MRI data was acquired on the 3T CONNECTOM system using a custom-built 64-channel array.  Data was acquired corresponding to a nominal 660 $\mu$m isotropic resolution over a 220$\times$118$\times$151.8 mm$^3$ FOV (matrix size 332$\times$180$\times$230), with 7 unweighted images (i.e., $b=0$) and 64 diffusion weighted images (DWIs) with $b=$1,500 s/mm$^2$ and uniformly distributed diffusion encoding orientations.  Data was measured using an EPI readout with 2 simultaneously-acquired thick slabs (i.e., each slab is 3.3 mm thick, which is 5 times the size of a 660 $\mu$m sub-slice) with blipped CAIPI \cite{setsompop2012} and sagittal slab orientations (matrix size for each slab volume was 332$\times$180$\times$46), 5 different RF encoding pulses, 6/8ths partial Fourier and 2$\times$ GRAPPA in-plane acceleration, ZOOPPA \cite{heidemann2012} outer volume suppression of the neck and phase-encoding along the superior-inferior axis, TE = 80~ms, and TR = 4.4~s per thick-slab volume (i.e., 22~s per DWI).  This entire measurement was acquired in $\approx$25 minutes, with three repetitions acquired to provide a gold standard reference (total $\approx$75 minutes).  Prior to further processing steps, slice-GRAPPA \cite{cauley2014} was used for SMS and parallel imaging reconstruction of the slab volumes, and in-plane registration was performed on these images using FSL \cite{jenkinson2012}.  

gSlider reconstruction was performed for single-average data using the conventional gSlider approach (i.e., phase estimation \cite{eichner2015} followed by regularized reconstruction according to Eq.~\eqref{eq:grecon}) and the SER approach described above.  A gold standard was obtained by applying conventional gSlider reconstruction (without SER) to each of the three repetitions, and the resulting phase-corrected real-valued images were then averaged together.

For comparison, we also applied denoising approaches based on low-rank matrix modeling to the images obtained from the conventional gSlider reconstruction.  Low-rank modeling has become a relatively popular approach for modeling and denoising diffusion MRI data in recent years  \cite{manjon2013,lam2013,haldar2011c,gao2014,veraart2016,ma2017}, and is based on the well-known denoising characteristics of principal component analysis (PCA).  We considered three different types of low-rank modeling.

First, following Ref.~\cite{veraart2016}, we applied sliding-window PCA denoising across the diffusion encoding dimension to overlapping spatial patches of size $12.5$mm $\times 12.5$mm $\times 12.5$mm, and chose the rank value for each patch based on the Marchenko-Pastur distribution. This approach, which we denote as MPPCA, is based on locally low-rank modeling assumptions. Data processing for this approach  was performed using the MRtrix3 package (\url{http://www.mrtrix.org/}).

While various methods exist for estimating the optimal parameters to use in low-rank matrix denoising \cite{veraart2016,candes2013,gavish2017} (including the previously-mentioned approach based on the Marchenko-Pastur distribution), we also were interested in understanding the best possible denoising performance that could be achieved using low-rank modeling constraints.  As a result, the next two low-rank modeling methods, which we call local PCA (LPCA) and global PCA (GPCA), both choose rank parameter values to minimize the mean-squared error of the denoised images with respect to the gold-standard images from $3\times$-averaged data.  Note that an explicit objective of Refs.~\cite{candes2013,gavish2017} is  to choose parameters that minimize the mean-squared error, but these methods make potentially suboptimal choices because they do not have access to the true mean-squared error values.  Since perfect adjustment of the rank parameter will not be available in practical experiments, the results obtained from LPCA and GPCA should be viewed as best-case scenarios for denoising with low-rank modeling constraints.   

While LPCA and GPCA both use the same approach for rank selection, they differ from one another in the way they choose image patch sizes.  Similar to MPPCA \cite{veraart2016}, LPCA applies sliding-window PCA denoising to overlapping patches of size $12.5$mm $\times 12.5$mm $\times 12.5$mm, following a locally low-rank data model.  Similar to Refs.~\cite{lam2013,haldar2011c}, GPCA applies PCA denoising to the whole 3D image volume at once, following a globally low-rank data model.

\subsection{Performance Metrics}

Denoising performance was assessed using several different metrics.  First, we computed the normalized root-mean-squared error (NRMSE) of the reconstructed DWIs with respect to the gold standard, defined as
\begin{equation}
NRMSE = \frac{\left\|\hat{\mathbf{f}} - \mathbf{f}_{\mathrm{gold}}\right\|_2}{\left\|\mathbf{f}_{\mathrm{gold}}\right\|_2}
\end{equation}
where $\mathbf{f}_{\mathrm{gold}}$ denotes the gold-standard result obtained from $3\times$ averaging.  While the gold-standard data is not entirely noise-free, it does have substantially better quality than the data acquired without averaging, so smaller values of this NRMSE metric are expected to correlate well with improved denoising performance.

For many practical applications, the errors in the denoised images are probably less important than errors in the quantitative diffusion parameters that can be estimated from the images.  As a result, we computed several different diffusion parameters.  Because diffusion data was collected with a single-shell q-space acquisition, we focused on quantitative parameters that can be estimated from single-shell data.

We started by using the software underlying the BrainSuite Diffusion Pipeline (\url{http://brainsuite.org/}) to estimate the diffusion tensor imaging (DTI) parameters of mean diffusivity (MD) and Fractional Anisotropy (FA).  To obtain quantitative error measures, we computed NRMSE values within the brain for both the FA and MD maps.  

We also used the software underlying the BrainSuite Diffusion Pipeline to estimate orientation distribution functions (ODFs) in two different ways: the Funk-Radon transform (FRT) \cite{tuch2004,hess2006,descoteaux2007} and the Funk-Radon and Cosine Transform (FRACT) \cite{haldar2013}.  The FRT and FRACT both estimate higher-order ODF representations  that, unlike the simpler DTI model, are capable of capturing complicated fiber-crossing structures in white matter.  Comparing these two methods, the FRT is expected to be less sensitive to noise than FRACT, while FRACT is expected to produce ODFs with higher angular resolution than the FRT.  Both of these methods are usually designed to be used with high b-value diffusion data sampled with a large number of diffusion encoding directions.  Since the data acquired in this work used a modest b-value with a modest number of encoding directions, the results of FRT and FRACT are not expected to be very impressive in this case, even for the gold standard data.  Nevertheless, we believe that the errors observed in the FRT and FRACT ODF estimates before and after denoising should still be reflective of denoising performance.  To obtain quantitative error measures, we computed NRMSE values within the white matter for the spherical harmonic coefficients used to represent the FRT and FRACT ODFs.  Specifically, NRMSE was computed over white matter voxels where the gold standard FA was \textgreater0.3.

\section{Results}

\begin{figure}[ht]
\includegraphics[width=\textwidth]{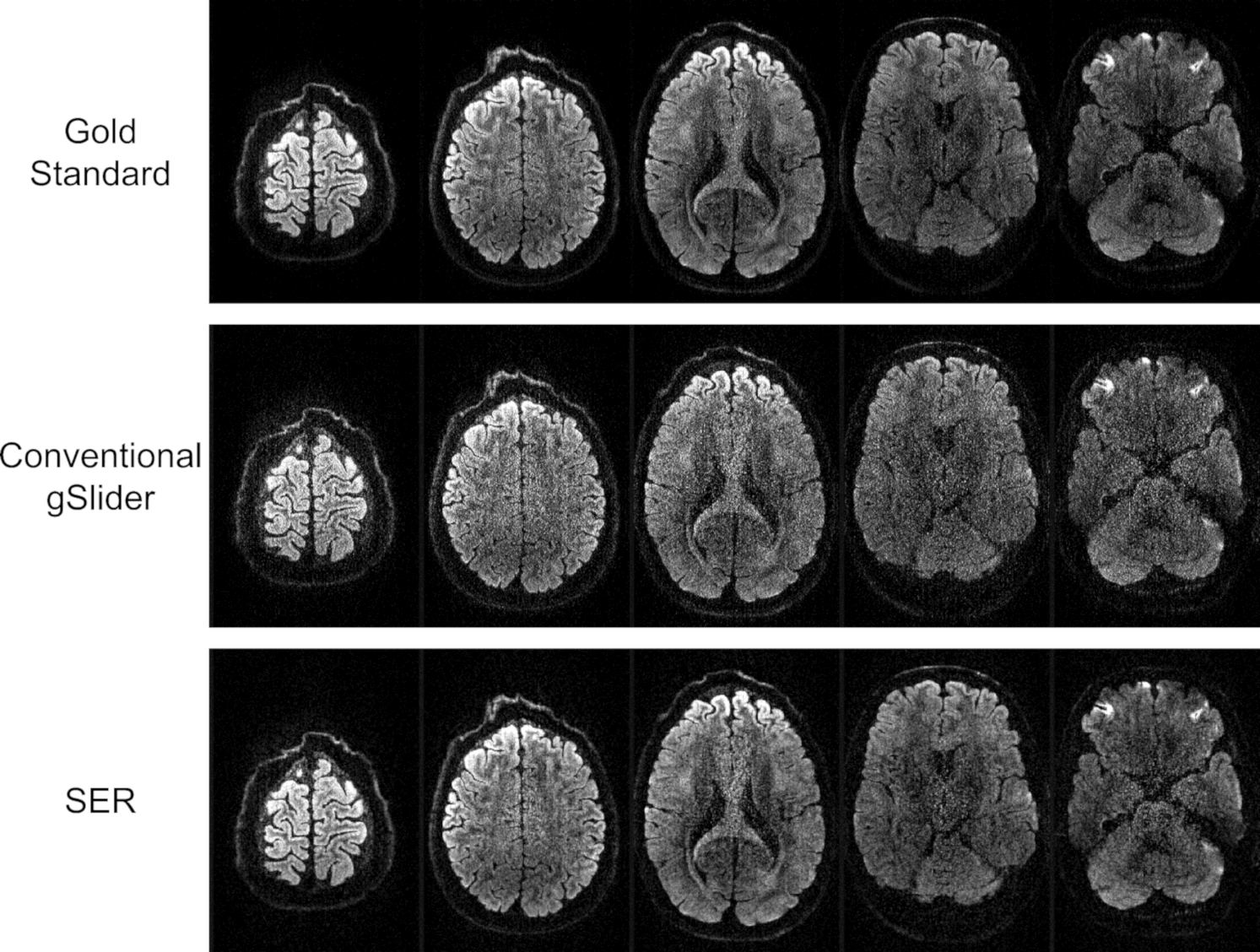}
\caption{Gold standard, conventional gSlider, and SER images from five slices of a representative DWI.}
\label{fig:dwi_slices}
\end{figure}

Gold standard, conventional gSlider, and SER images from several slices of a representative DWI are illustrated in Fig.~\ref{fig:dwi_slices}.  As can be seen, the conventional gSlider result has a noisy appearance, particularly near the center of the brain.  On the other hand, the SER has a much less noisy appearance, as should be expected.  

\begin{figure}[ht]
\includegraphics[width=\textwidth]{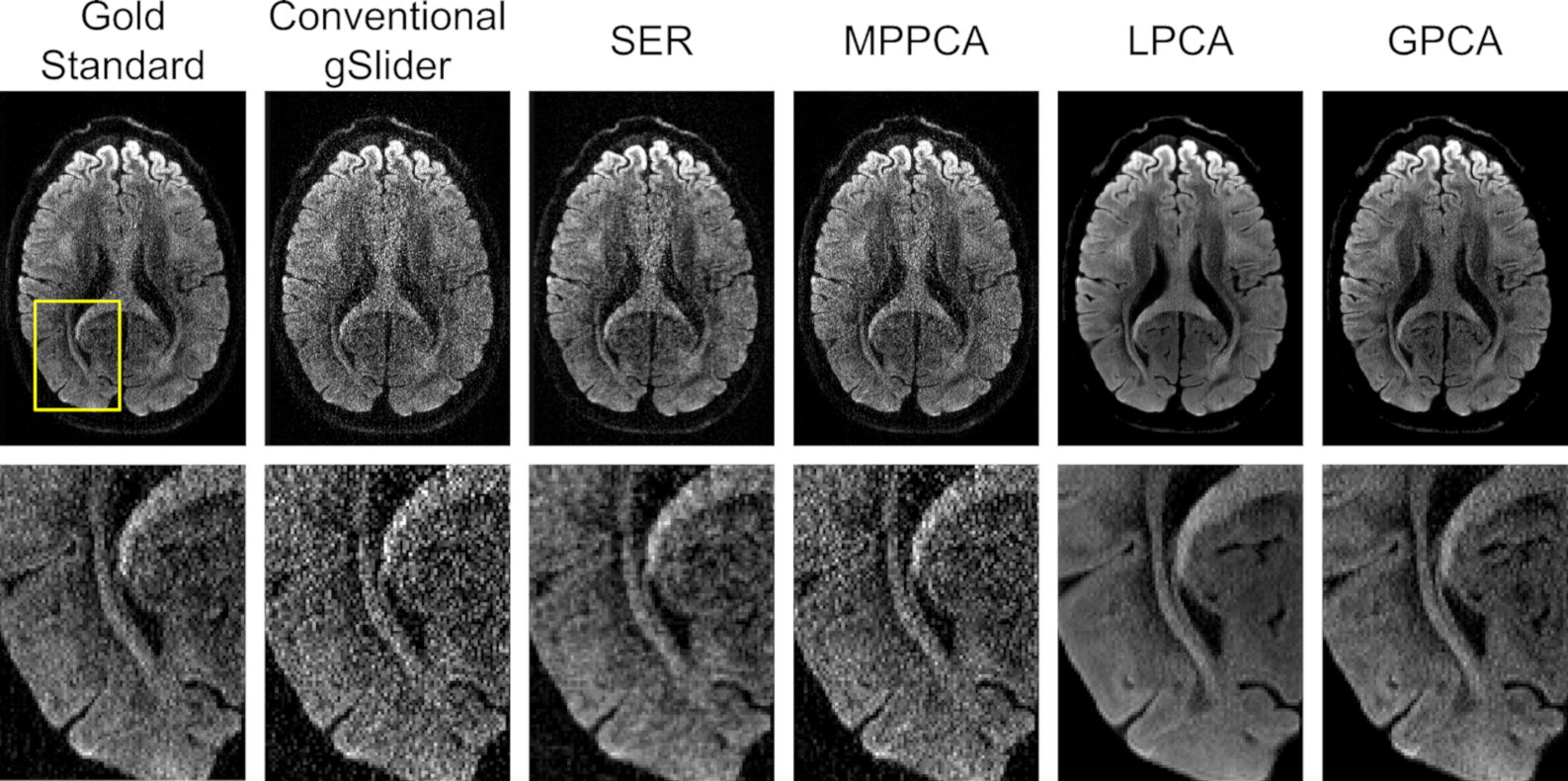}
\caption{Comparison of a slice from a representative DWI obtained by different reconstruction and denoising approaches.  Zoomed-in images (from the region corresponding to the yellow box shown in the gold standard image) are also shown for easier visualization of fine image details.}
\label{fig:dwi_comparisons}
\end{figure}

More detail is available from inspecting  the zoomed-in images presented in Fig.~\ref{fig:dwi_comparisons}, which also includes comparisons against MPPCA, LPCA, and GPCA.  From this figure, it can be observed that all four denoising approaches  have an apparent reduction in noise, although the extent of this varies from method to method.  MPPCA appears to have the least amount of noise reduction among the four, while both GPCA and LPCA produce very crisp-looking images with minimal obvious noise content.  We would argue that the SER result does not look as crisp or cosmetically pleasing as the LPCA and GPCA results, although still represents an improvement over MPPCA.  

Looking at the NRMSE values for the DWIs, which are shown in Table~\ref{tab:err}, we observe that LPCA has the smallest NRMSE, followed by SER, GPCA, and MPPCA.  The good performance of LPCA in this case is not surprising, since its parameters were tuned using perfect knowledge of the gold standard to optimize this NRMSE value, and because LPCA has more flexibility than GPCA to adapt to the different properties of different image regions.  On the other hand, it may be surprising that the SER result has smaller NRMSE than the GPCA result, given that the GPCA result was also designed to minimize this NRMSE value based on perfect prior knowledge that was not available to SER.  This suggests that even in the ideal case, while global low-rank modeling can offer some degree of improvement,  this model may be intrinsically more limited than some of these other approaches in this type of application.  Interestingly, there is a major performance difference between LPCA and MPPCA, even though both of these results are based on the same underlying locally low-rank data model.  This result illustrates the importance of choosing good rank parameters, although it should be noted that MPPCA uses a conservative automatic rank-selection rule that is not necessarily designed to achieve optimal NRMSE.

\begin{figure}[ht]
\includegraphics[width=\textwidth]{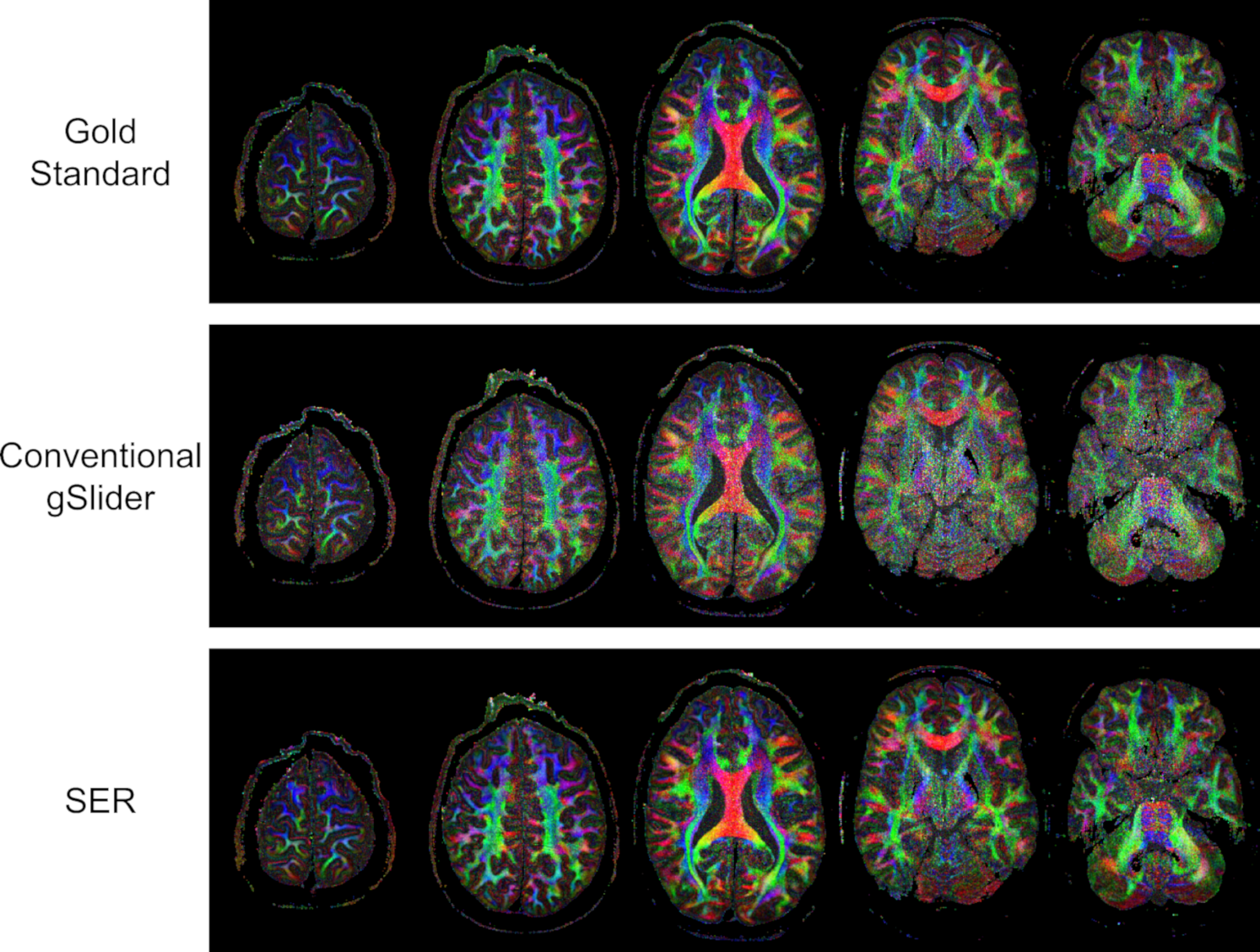}
\caption{Color FA maps computed based on gold standard, conventional gSlider, and SER images from five slices of the brain volume.}
\label{fig:dti_slices}
\end{figure}

\begin{figure}[ht]
\includegraphics[width=\textwidth]{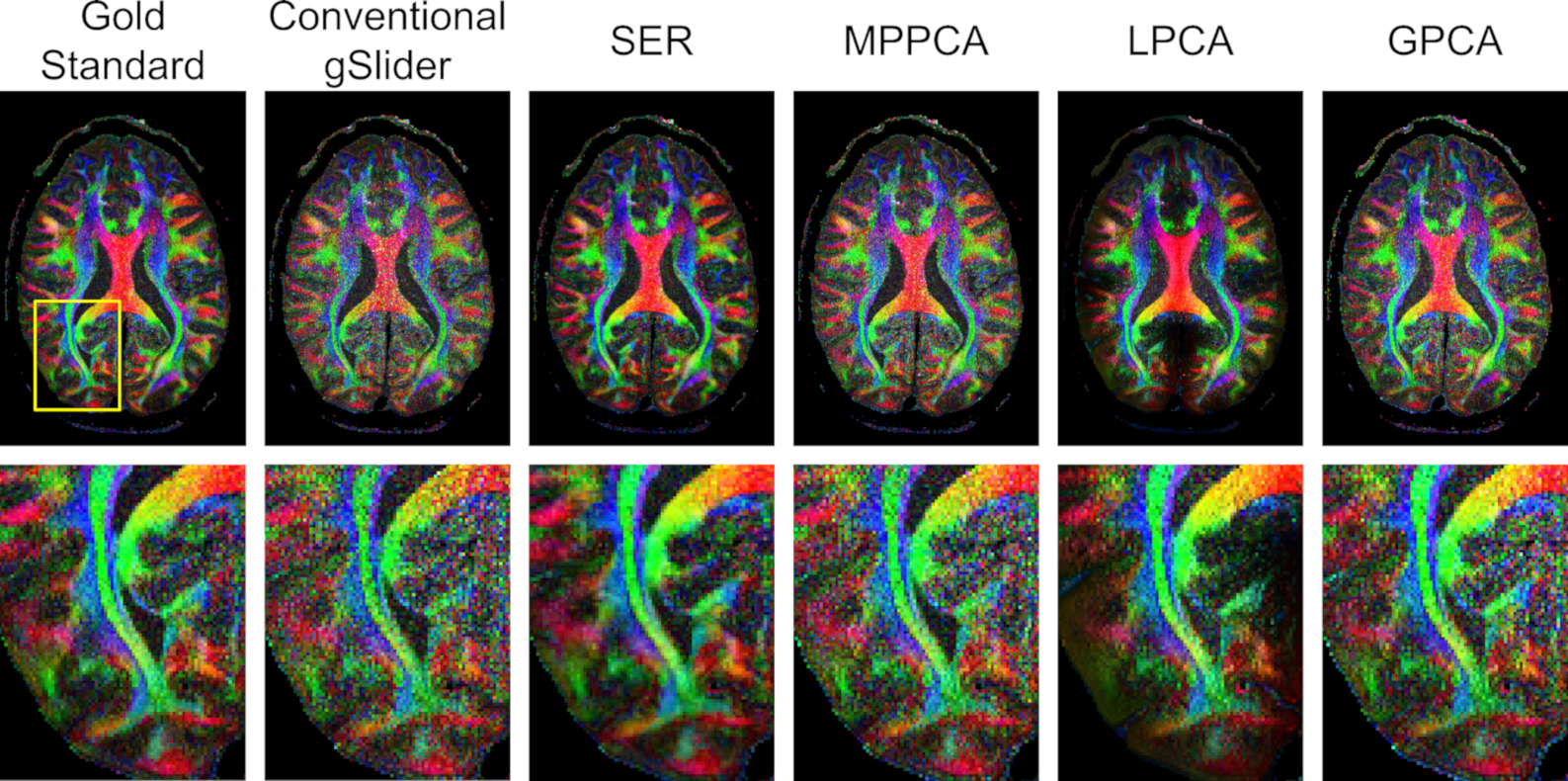}
\caption{Comparison of color FA maps obtained by different reconstruction and denoising approaches. Zoomed-in images (from the region corresponding to the yellow box shown in the gold standard map) are also shown for easier visualization of fine details. }
\label{fig:dti_comparisons}
\end{figure}

Figures~\ref{fig:dti_slices} and \ref{fig:dti_comparisons} respectively show color-coded FA maps corresponding to the same images from Figs.~\ref{fig:dwi_slices} and \ref{fig:dwi_comparisons}, and NRMSE values corresponding to the quantitative DTI parameters (MD and FA) are also shown in Table~\ref{tab:err}.  Interestingly, SER and MPPCA are now the two best denoising methods with respect to these DTI parameters, and SER has  a substantial advantage over MPPCA.  Visually, Fig.~\ref{fig:dti_comparisons} shows that SER results in a color FA map that does not appear to be as noisy as the color FA maps from the other methods.   

Surprisingly, even though LPCA had the smallest NRMSE with respect to the DWIs, it actually has the largest NRMSE for these quantitative DTI parameters.  Even worse, the NRMSE values in these cases are even substantially larger than those obtained from  conventional gSlider without any denoising.  This likely occurs because quantitative model-fitting in diffusion MRI can be sensitive to small variations in the measured diffusion signal, while by its nature, low-rank modeling seeks to find a low-dimensional subspace representation that preserves signal energy as much as possible, which does not prioritize the preservation of smaller signal variations.  This result also underscores the more general point that visually pleasant images with small error values will not necessarily yield good results when those images are used in quantitative applications, a concern that has also been raised in previous related work \cite{haldar2012,haldar2011c,haldar2008,lam2013}.    

Another interesting observation is that the color FA map for  GPCA from Fig.~\ref{fig:dti_comparisons} has a much noisier-looking appearance than might have been expected based on the relatively noise-free appearance of the corresponding DWIs from Fig.~\ref{fig:dwi_comparisons}.  This might be explained by the fact that the DTI model is itself low-dimensional, such that the space of all possible signals that are compatible with the DTI model may already   approximately reside in a low-dimensional subspace \cite{lam2013}.  Thus, the DTI fitting procedure  may have an effect that is similar in some sense to low-rank modelling, and we might not expect low-rank denoising to yield a dramatic improvement in DTI parameters unless the explicit subspace constraints imposed by low-rank modeling are sufficiently distinct from the implicit subspace constraints associated with DTI modeling.  In the case of global low-rank modeling (as used by GPCA), the subspace constraints are required to apply broadly to the data from all voxels, and so may not be very distinct from the implicit subspace constraints imposed by the DTI model.  

\begin{figure}[ht]
\includegraphics[width=\textwidth]{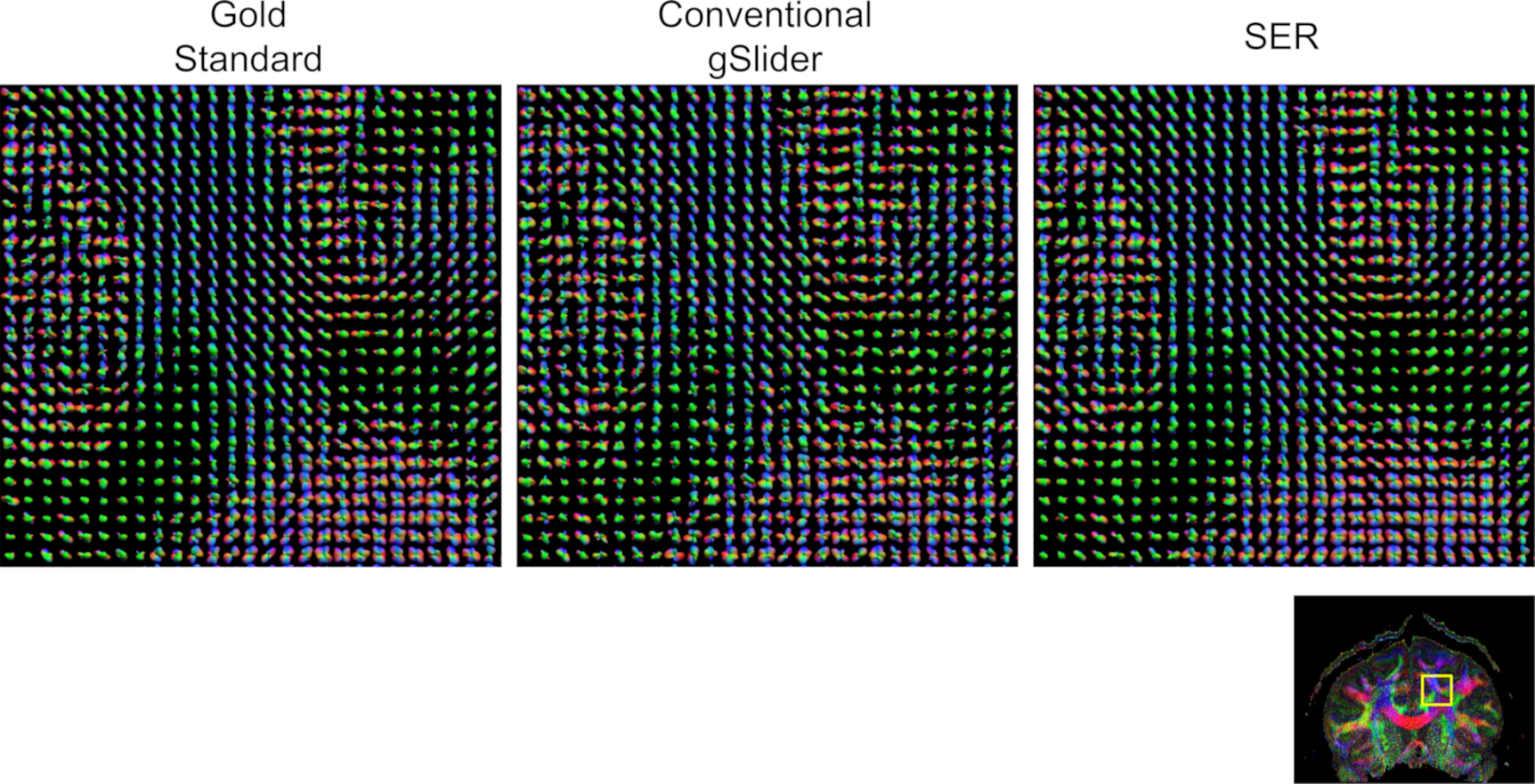}
\caption{ODFs estimated using the FRT from gold standard, conventional gSlider, and SER images.  ODFs are shown from the brain region indicated by the yellow box in the reference color FA image. }
\label{fig:odf_frt}
\end{figure}

\begin{figure}[ht]
\includegraphics[width=\textwidth]{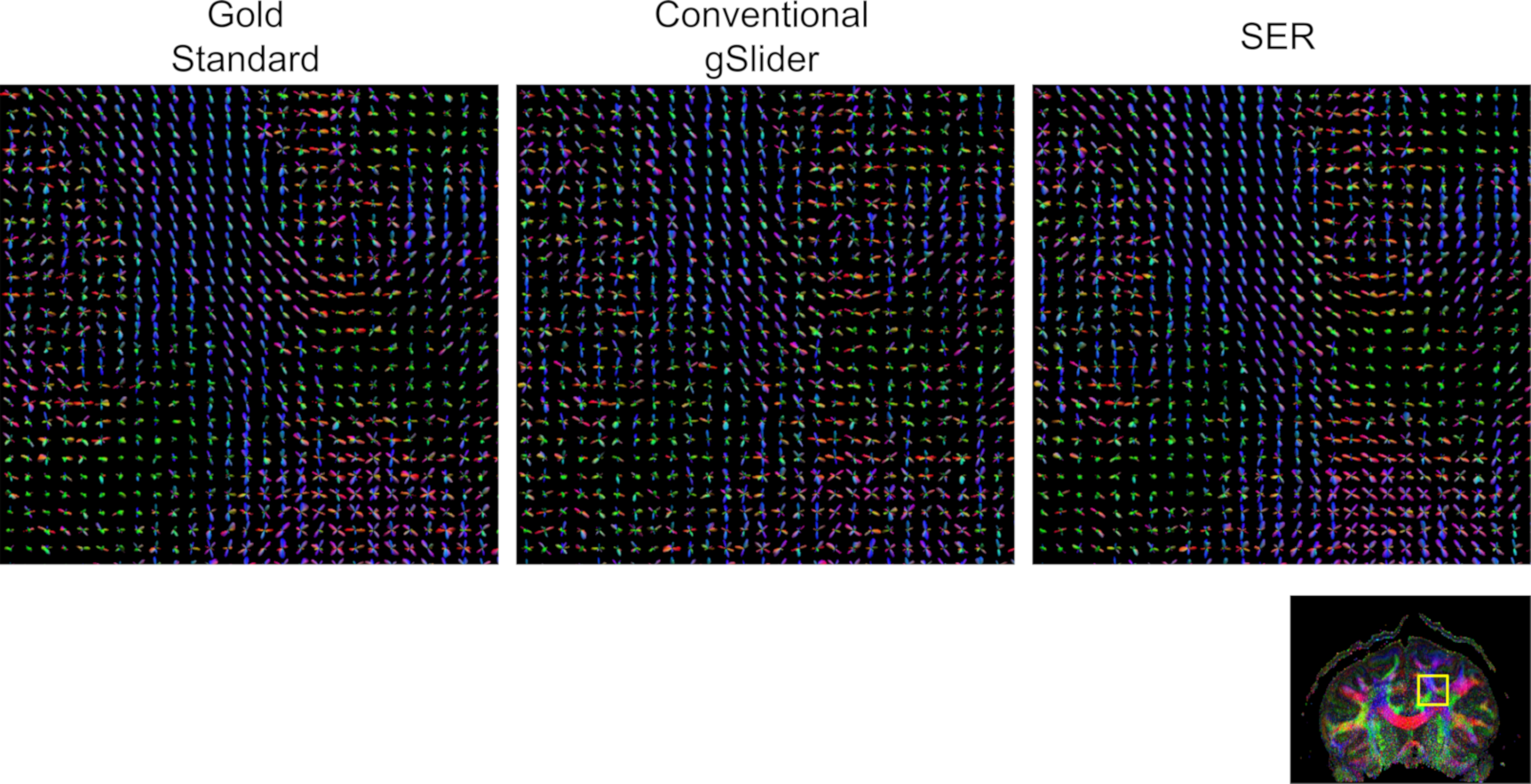}
\caption{ODFs estimated using the FRACT from gold standard, conventional gSlider, and SER images.  ODFs are shown from the brain region indicated by the yellow box in the reference color FA image.}
\label{fig:odf_fract}
\end{figure}

Figures~\ref{fig:odf_frt} and \ref{fig:odf_fract} respectively show ODF estimation results for FRT and FRACT, with quantitative NRMSE values also reported in Table~\ref{tab:err}.  The figures show that the spatial distribution of ODFs appears somewhat chaotic and disorganized when ODFs are estimated from the noisy conventional gSlider images, while the ODFs estimated from SER have more spatial coherence and match better with the characteristics of the ODFs estimated from the gold standard images.  Quantitatively, SER still possesses the smallest NRMSE values for both FRT and FRACT ODFs.  LPCA and MPPCA are virtually tied for second place with respect to FRT ODFs, and LPCA stands alone in second place for FRACT ODFs.

In addition to achieving good quantitative performance, SER also has the advantage that it is possible to theoretically characterize its noise and resolution characteristics \cite{haldar2011c,haldar2012}.  This allows us to have a deeper understanding of the limitations and trade-offs associated with SER reconstruction.  While MPPCA, LPCA, and GPCA certainly should also be associated with various limitations and trade-offs, we are unaware of any convenient methods to compute characterizations of these trade-offs.

\begin{figure}[ht]
\includegraphics[width=\textwidth]{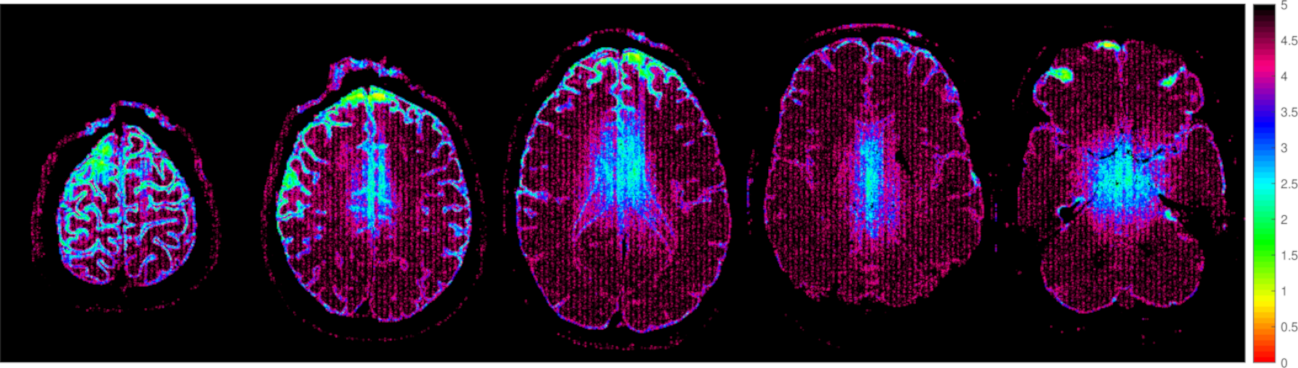}
\caption{Spatial maps of the expected reduction in noise variance obtained by using SER instead of conventional gSlider reconstruction.  Maps are shown corresponding to the same image slices from Figs.~\ref{fig:dwi_slices} and \ref{fig:dti_slices}.}
\label{fig:snr}
\end{figure}

Figure~\ref{fig:snr} shows spatial maps of the expected reduction in noise variance obtained by using SER instead of conventional gSlider reconstruction, which were obtained using the exact same techniques described in previous SER publications\cite{haldar2011c,haldar2012}.    As noted previously, the regularization parameters for SER were designed to achieve at least a 3$\times$ reduction in noise variance across most brain regions.  For reference, note that an expected 3$\times$ reduction in noise variance is commensurate with 3$\times$ averaging.  As can be seen from this figure, the reduction in noise variance varies spatially, with larger improvements in SNR in smooth image regions, and smaller improvements in SNR near estimated image edge locations.  This behavior is expected, and results from the fact that SER is designed to avoid smoothing across edge structures to preserve high-resolution image content\cite{haldar2012,haldar2011c}.  Notably,  central regions of the brain are also observed to have less SNR-enhancement than outer portions of the brain, even though these regions are not necessarily associated with strong edge content.  This occurs because the center of the brain has lower SNR than brain regions with closer proximity to the receiver coils, and SER has a more difficult time identifying the difference between actual image edge structure and noise in this region.  Overall, the level of noise variance reduction ranged from approximately $3$-$5$ across the entire image.  

\begin{figure}[ht]
\includegraphics[width=\textwidth]{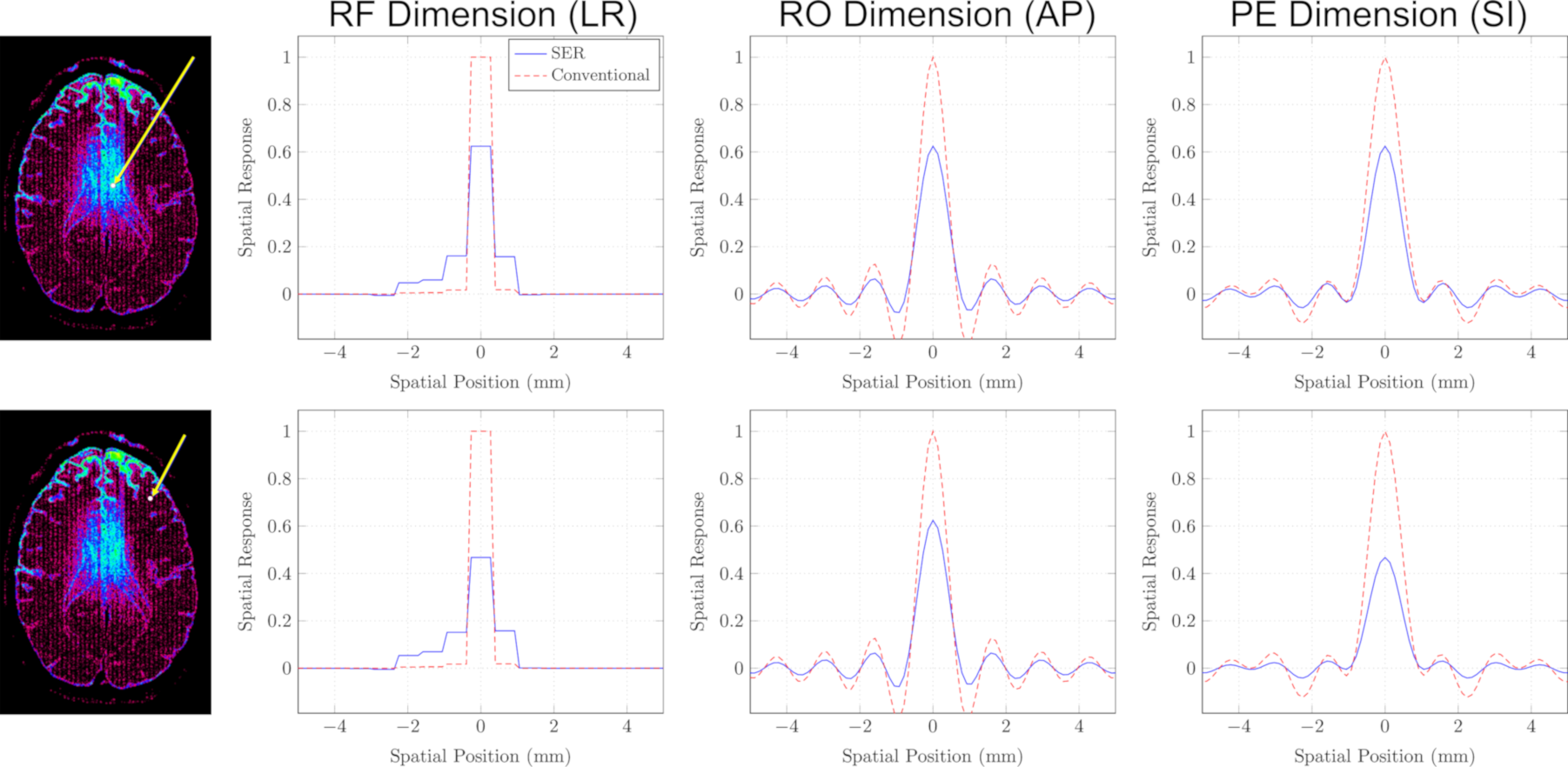}
\caption{SRFs for SER and conventional gSlider obtained from (top row) a voxel where the SNR improvement associated with SER was approximately 3 and (bottom row) a voxel where the SNR improvement associated with SER was approximately 5. The voxel positions are indicated as shown in the images on the left.  The SRFs in this case are three-dimensional functions, which are hard to display.  For easier visualization, we have shown  one-dimensional plots passing through the peak of the SRF along different orientations.  Specifically, we show SRF plots along the RF-encoding dimension (left-right anatomically), the readout encoding dimension (anterior-posterior anatomically), and the phase encoding dimension (superior-inferior anatomically).}
\label{fig:resolution}
\end{figure}

Importantly, even though SER employs edge-preserving regularization, the SNR improvement associated with SER is also associated with a degradation in spatial resolution, and this degradation can be characterized theoretically using point-spread functions/spatial response functions (SRFs)\cite{haldar2011c,haldar2012}. Using techniques described in previous publications\cite{haldar2011c,haldar2012,haldar2009c}, we calculated SRFs for one voxel where the noise variance reduction was approximately 3 and another voxel where the noise variance reduction was approximately 5, with results shown in Fig.~\ref{fig:resolution}.  For reference, we also show SRFs for conventional gSlider on the same axes.  In both cases, the use of SER is associated with a small degradation in spatial resolution.   

While there is no unique way of measuring spatial resolution \cite{haldar2011a}, we have opted to use the full-volume at half-maximum (FVHM) of the SRF as a resolution measure.  The FVHM is calculated by summing the volumes of all voxels for which the SRF is larger than half of its maximum value, and can be viewed as a generalization of the conventional full-width at half maximum for one-dimensional point-spread functions to the three-dimensional setting.  We observe that the FVHM of the SRF is fairly shift-invariant for conventional gSlider (as we had expected), with a value of roughly 0.380 mm$^3$ = (724 $\mu$m)$^3$ for all voxels.  In contrast, the FVHMs for SER vary with the level of SNR improvement.  For the voxel with $3\times$ noise variance reduction, the FVHM of the SRF for SER was roughly 0.426 mm$^3$ = (752 $\mu$m)$^3$, while the FVHM of the SRF was roughly 0.472 mm$^3$ = (778 $\mu$m)$^3$ for the voxel with $5\times$ noise variance reduction.   Consistent with previous SER publications\cite{haldar2012,haldar2008,haldar2011c,kim2015a,haldar2009c,haldar2008a,haldar2011a} (and perhaps suprising from the perspective of conventional imaging expectations), the SNR improvement associated with SER is quite substantial compared to the relatively minor degradations observed in spatial resolution.  In addition, it should also be remembered that SER is designed to avoid the blurring of information across image edges.  In particular, the SRF for SER adapts itself to avoid signal from leakaging across the edge structures of the image \cite{haldar2008,haldar2011c}, such that even this minor loss in resolution may not be as deleterious as it might have been otherwise.  

This figure also reveals that the spatial resolution characteristics for both SER and conventional gSlider are different along the RF-encoding dimension than they are along either of the two Fourier-encoding dimensions, as should be expected due to the use of substantially different spatial encoding mechanisms.  In addition, we observe substantial asymmetry in the SRF along the RF-encoding dimension.  While we have only shown results from two voxels, results from other voxels reveals that the shape of the SRF varies considerably depending on the position of the reconstructed voxel within the RF-encoded thick-slab that was used for acquisition.  Reconstructed voxels from the center of a thick slab have SRFs that are more symmetric than voxels that are closer to the edges of the thick slabs.   We also observe the tendency to smooth voxels more within the same slab rather than across different slabs. This is likely related to the structure of the inverse problem, where there is more ambiguity within the same slab than there is across slabs.   The effects of this are also obsereved in the noise variance reduction maps from Fig.~\ref{fig:snr}, where some striping is observed along slab boundaries.

\section{Discussion}

Overall, while SER did not have the smallest NRMSE values with respect to the denoised DWIs, it did have the smallest NRMSE values in all other cases compared to globally and locally low-rank modeling methods.  In addition, SER has useful theoretical characterizations that are generally not available for low-rank denoising approaches.  While we cannot comprehensively compare SER against all existing diffusion MRI denoising methods, we should note that previous work has compared SER against non-local means denoising with similarly promising results \cite{haldar2012,haldar2011c,kim2015a}.

While this paper presented a comparison of SER against low-rank modeling methods, it should be noted that SER and low-rank modeling are based on different principles and can potentially be combined synergistically.  It has been previously demonstrated in a different context that combining SER with globally low-rank modeling can lead to further performance improvements \cite{lam2013}. Based on our results, we expect that the combination of SER with locally low-rank modeling could be even better.  However, one challenge is that the use of any low-rank modeling can invalidate the theoretical characterizations of SER, while the use of SER can invalidate the noise modeling assumptions of automatic rank-selection rules.  This can cause parameter selection for both approaches to become more difficult.

Slow computation speed is one of the limitations of our current implementation, and for example, the results shown in this paper took multiple days to compute using a simple unoptimized Matlab implementation running on a Linux-based workstation with two quad-core Xeon 2.27 GHz processors and 48 GB of RAM.  Most of this time (\textgreater90\%) was associated with solving the phase-update step associated with Eq.~\eqref{eq:phi}, although the phase estimation is expected to be important as illustrated in Supporting Information Fig.~\ref{fig:phase}.    While slow computation speed may be understandable given the very large size of this dataset (i.e., the gSlider data vector $\mathbf{b}$ alone requires more than 15GB of memory to store in its entirety in double precision!), it would of course be preferable if this computation were faster.  Our group has recently been exploring faster algorithms for this type of optimization problem that have the potential to substantially reduce the amount of computation time \cite{liu2019a}, although we believe that a thorough exploration of different algorithmic alternatives is beyond the scope of the present paper.

In the implementation of SER that was used for this paper, we relied on complex-valued images obtained from an initial parallel imaging and SMS reconstruction procedure, instead of performing SER from raw k-space data as advocated in previous SER papers \cite{haldar2012,haldar2011c}.  This choice was made because the the raw k-space data for this acquisition was too big for easy manipulation and storage, and even if the k-space data had been stored, SER reconstruction from raw k-space data would have been practically unworkable without relying on specialized computional resources.  As a result, the SER results were obtained from data possessing less information content than the original raw data, and for example, we did not have access to the complementary information provided by different channels or information about the spatially-varying noise characteristics associated with the images.  We anticipate that denoising quality could potentially have been much better if at least a little more of the original information had been preserved.  For example, if information about the spatially-varying noise variance had been preserved, it may have been possible to use this information to avoid the situation where the center of the brain experiences less denoising than exterior parts of the brain.  While there exist various techniques for trying to estimate the characteristics of spatially-varying noise fields as an inverse problem \cite{ajafernandez2016},  there are also direct ways of precisely calculating and preserving this information at the time of image reconstruction \cite{varadarajan2015,robson2008}, and we believe that exploring the use of such noise maps is a promising direction for future work.

One of our observations from this study was that basic SER reconstruction yields different resolution characteristics along the RF-encoded dimension than for the Fourier-encoded dimensions.  It is not immediately clear whether this difference in resolution characteristics is problematic.  However, if it is ultimately deemed to be problematic, it is interesting to note that technqiues  already exist for modifying regularized reconstruction methods to achieve more uniform resolution characteristics \cite{ahn2008,stayman2000}.  The application of such ideas to SER may also be an interesting topic for future research.

Finally, we should note that the results shown in this paper are based on the use of full RF-encoding for every DWI in acquisition together with spatial smoothness constraints for SNR-enhancement, but without any constraints on the structure of the data in q-space.  In principal, several groups have previously shown that if a diffusion MRI acquisition requires multiple encodings for each DWI, it can be possible to eliminate certain encodings by exploiting the smoothness of the q-space signal\cite{bhushan2014,ning2016,vansteenkiste2016}. This approach is complementary to SER, and a preliminary exploration of combining SER   with subsampled RF encoding has shown promising results \cite{haldar2017b}.  Further exploration of this kind of approach may enable even faster high-resolution diffusion MRI using gSlider-SMS.

\section{Conclusion}

This work proposed a new approach to fast diffusion MRI that uses a highly-efficient acquisition strategy together with an advanced and theoretically-characterizable denoising strategy to enable a relatively fast diffusion MRI experiment with state-of-the-art volume coverage, spatial resolution, and SNR.  We believe that this kind of approach can prove useful across the full range of in vivo human diffusion MRI applications.
 
\providecommand{\noopsort}[1]{}

\clearpage 
\section*{Figure and Table Captions}

\clearpage 

\clearpage 

\clearpage 

\clearpage 

\clearpage 

\clearpage 

\clearpage 

\clearpage 

\clearpage 
\begin{table}[ht]
\includegraphics[width=\textwidth]{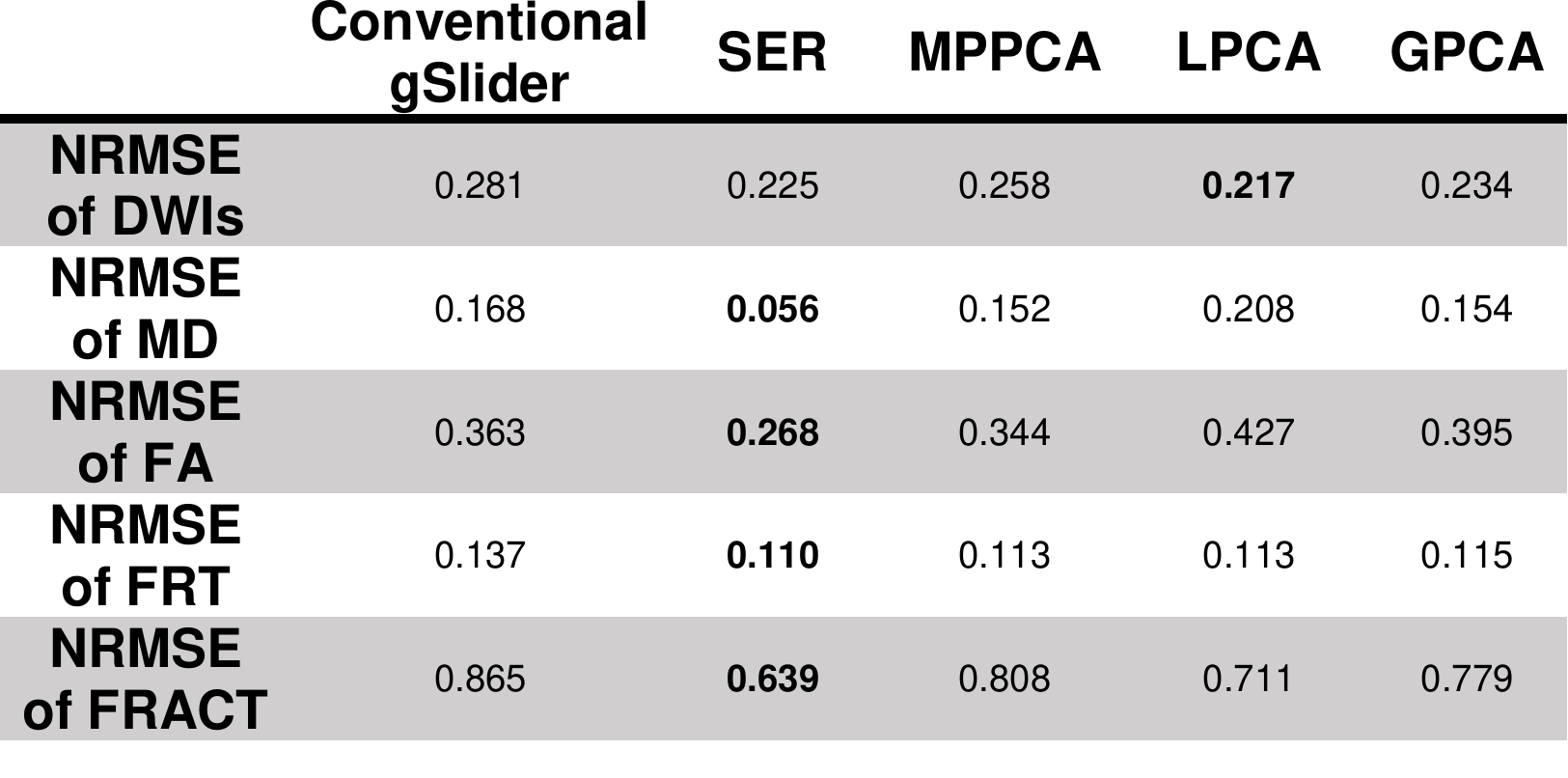}
\caption{Quantitative error measures obtained with different reconstruction and denoing methods.}
\label{tab:err}
\end{table}

\clearpage
\setcounter{figure}{0} 
\renewcommand{\figurename}{Supporting Information Figure}
\renewcommand{\thefigure}{S\arabic{figure}}

\begin{figure}[ht]
\includegraphics[width=\textwidth]{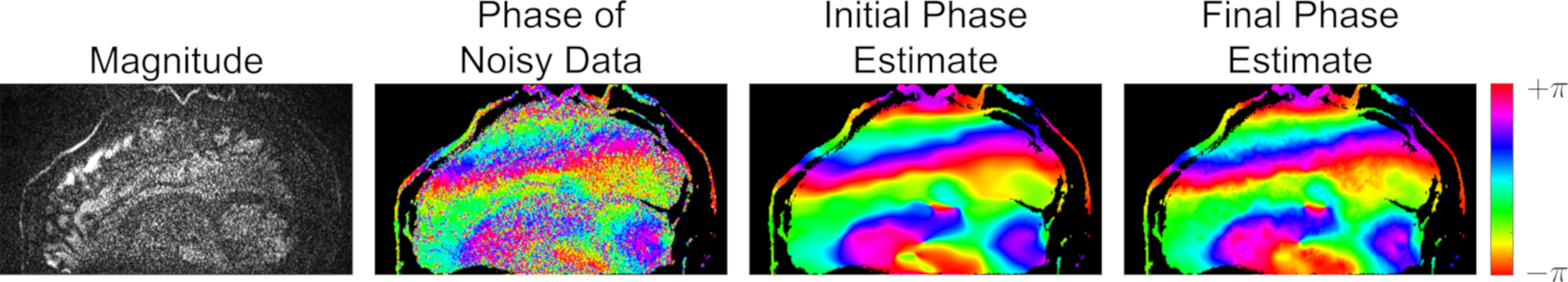}
\caption{Illustration of the effects of iterative regularized phase estimation.  As can be seen, the initial phase estimate obtained from a low-resolution reconstruction of the image has much less noise than the original image, though the resulting phase map is potentially oversmoothed.  In contrast, the final phase estimate obtained from iterative regularized phase estimation is able to capture higher-resolution spatial phase variations.}
\label{fig:phase}
\end{figure}

\end{document}